\documentclass{article}

{}
{}
{}

\usepackage{amsmath} 
\usepackage{amsthm} 
\usepackage{cite} 
\usepackage{hyperref} 
\usepackage{graphicx}
\usepackage{algorithmic} 
\usepackage{hyperref}
\usepackage{authblk}
\usepackage[margin=1.2cm]{geometry}
\usepackage{multicol}
\usepackage{blindtext}
\usepackage{epstopdf}
\usepackage{indentfirst}
\usepackage{caption}
\begin{document}

\title{Empowering Aggregators with Practical Data-Driven Tools: Harnessing Aggregated and Disaggregated Flexibility for Demand Response}

\author[1]{Costas Mylonas}
\author[2]{Donata Borić}
\author[2]{Leila Luttenberger Marić}
\author[3]{Alexandros Tsitsanis}
\author[1]{Eleftheria Petrianou}
\author[1]{Magda Foti}

\affil[1]{{UBITECH, Thessalias 8, 15231, Athens, Greece,} E-mail: kmylonas@ubitech.eu, epetrianou@ubitech.eu, mfoti@ubitech.eu \newline}
\affil[2]{{KONČAR-Digital Ltd. for Digital Services, Fallerovo šetalište 22, HR-100000, Zagreb, Croatia} E-mail: donata.boric@koncar.hr, leila.luttenberger@koncar.hr \newline}
\affil[3]{{Suite5 Data Intelligence Solutions, Alexandreias 2, 3013, Limassol, Cyprus} E-mail: alexandros@suite5.eu \newline}
\date{}

\setcounter{Maxaffil}{0}
\renewcommand\Affilfont{\itshape\small}

\maketitle

\thanks{This paper is a preprint of a paper submitted to the 14th Mediterranean Conference on Power Generation, Transmission, Distribution and Energy Conversion (MEDPOWER 2024) and is subject to Institution of Engineering and Technology Copyright. If accepted, the copy of record will be available at IET Digital Library.}

\begin{abstract}
\noindent This study explores the interaction between aggregators and building occupants in activating flexibility through Demand Response (DR) programs, with a focus on reinforcing the resilience of the energy system considering the uncertainties presented by Renewable Energy Sources (RES). Firstly, it introduces a methodology of optimizing aggregated flexibility provision strategies in environments with limited data, utilizing Discrete Fourier Transformation (DFT) and clustering techniques to identify building occupants' activity patterns. Secondly, the study assesses the disaggregated flexibility provision of Heating Ventilation and Air Conditioning (HVAC) systems during DR events, employing machine learning and optimization techniques for precise, device-level analysis. The first approach offers a non-intrusive pathway for aggregators to provide flexibility services in environments of a single smart meter for the whole building's consumption, while the second approach maximizes the amount of flexibility in the case of dedicated metering devices to the HVAC systems by carefully considering building occupants' thermal comfort profiles. Through the application of data-driven techniques and encompassing case studies from both industrial and residential buildings, this paper not only unveils pivotal opportunities for aggregators in the balancing and emerging flexibility markets but also successfully develops and demonstrates end-to-end practical tools for aggregators.
\end{abstract}

\begin{multicols}{2}

\section{Introduction}

Consumer engagement in Demand Response (DR) programs plays a pivotal role in harnessing flexibility within the energy system, essential for achieving decarbonization goals and maintaining system resilience. Aggregators serve as key intermediaries, coordinating between consumers and the energy market to optimize demand-side flexibility. Their role is increasingly recognized as vital in the European Union's regulatory framework, which emphasizes the importance of aggregators in managing distributed energy resources (DER) and integrating renewable energy sources (RES) into the grid. The latest European policies highlight the need for aggregators to enable the participation of small-scale consumers in the electricity market, thus ensuring a more reliable and sustainable energy system \cite{kerscher2022key}.

Aggregated and disaggregated flexibility provision are critical strategies used by aggregators to manage demand-side resources. Aggregated flexibility involves managing energy loads at the building or multi-building level, optimizing overall consumption without focusing on individual devices. This broad approach is typically used when only whole-building data is available. In contrast, disaggregated flexibility targets specific devices within a building, utilizing detailed data from sub-metering to optimize individual systems like Heating Ventilation and Air Conditioning (HVAC) or heat pumps. Authors of \cite{bachoumis2023data} emphasize the significance of these approaches in the Buildings-to-Grid (B2G) framework, where adaptive strategies and data-driven analytics are key to ensuring grid reliability and efficient demand response.

Aggregated flexibility provision at the building or multi-building level involves managing the collective energy consumption of buildings to optimize overall load and contribute to DR programs. This approach is particularly useful when detailed, device-level data is not available, making it necessary to work with aggregated data from entire buildings. For example, in the context of day-ahead load forecasting, methods such as deep learning and unsupervised clustering have been employed to predict the energy consumption patterns of multiple households, allowing aggregators to optimize energy management strategies at a broader level \cite{budin2022day}. Similarly, the potential of demand-side resources across residential and commercial buildings to provide flexibility has been explored, highlighting the role of aggregators in harnessing these resources for grid stability \cite{anwar2018harnessing}. Techniques like neural networks and optimization algorithms have also been applied to forecast and manage demand-side flexibility using aggregated building-level data, proving effective in scenarios where only overall consumption data is available \cite{merce2020load}. These methodologies enable aggregators to implement broad DR strategies, making the most of available data to meet the requirements of both the energy market and regulatory frameworks \cite{goldberg2013measurement}.

Disaggregated flexibility provision focuses on managing energy consumption at the device level, enabling more granular and precise DR strategies. This approach is particularly important for optimizing the performance of specific devices, such as HVAC systems, within buildings. Several studies emphasize the significance of device-level management in enhancing the overall flexibility of the energy system. For instance, the study on the control of aggregated inverter air conditioners \cite{che2019demand} demonstrates how individual and collective control of air conditioning units can provide significant flexibility by adjusting the load based on real-time demand. Similarly, study \cite{alic2020consumer} discuss consumer-driven energy management strategies for HVAC systems, focusing on the potential of device-level interventions to optimize energy use without compromising comfort. Further, the quantification of flexibility at the device level has been shown to be particularly significant for HVAC systems, where precise control mechanisms are implemented to optimize demand response outcomes \cite{chen2019quantification}. Techniques for estimating HVAC flexibility under various conditions further support the ability of these systems to contribute effectively to DR programs, highlighting the importance of detailed, device-level analysis \cite{yin2016quantifying}.

Beyond HVAC systems, the integration of privacy-preserving algorithms, optimization techniques, and machine learning models has enabled advanced management of device-level flexibility across various appliances. The optimization of appliance scheduling at the device level, particularly through machine learning techniques, has been instrumental in refining the contributions of individual devices to overall demand response strategies \cite{song2018cluster}. Moreover, privacy-preserving disaggregation methods allow for the non-intrusive management of flexible energy resources, ensuring that individual consumption profiles are protected while still enabling effective DR participation \cite{jacquot2019privacy}. In a related context, study \cite{foti2023privacy} explores a privacy-preserving transactive energy system using homomorphic encryption, which protects participant privacy while enabling real-time energy market transactions through a uniform price double auction mechanism. Such privacy-preserving approaches are crucial for gaining consumer trust and facilitating broader adoption of smart grid technologies.These approaches are further supported by machine learning models designed to predict and manage device-level flexibility, enhancing the ability of aggregators to optimize DR programs \cite{danner2021flexibility}.

This paper introduces an integrated toolkit for flexibility provision that effectively combines both aggregated and disaggregated strategies, addressing a gap in the existing literature where these approaches are often considered separately. By applying advanced data analytics such as Discrete Fourier Transformation (DFT) and clustering techniques, our method optimizes flexibility at the building level, making it effective even when only aggregated data is available. Simultaneously, our focus on disaggregated flexibility, particularly for HVAC systems, leverages machine learning and optimization techniques to ensure precise control while maintaining occupant thermal comfort, which is crucial for engagement in DR programs. Unlike prior studies that concentrate solely on either broad building-level strategies or detailed device-level interventions, this research uniquely integrates both, offering a practical, end-to-end solution designed specifically for aggregators. The development of a comprehensive application further underscores the applicability of our method, providing aggregators with a robust toolkit to optimize demand-side flexibility across diverse building environments.

\section{Aggregated Flexibility Provision}

The optimization of aggregated flexibility provision strategies begins with accurately identifying recurring energy consumption patterns, such as those occurring on an hourly, daily, or weekly basis. To do this, we employ DFT, a technique that transforms time-series data into the frequency domain. This transformation helps us to pinpoint the dominant frequencies or cycles in the data, such as a daily peak in energy use. By understanding these recurring patterns, we can then apply the Density-Based Spatial Clustering of Applications with Noise (DBSCAN) algorithm to group similar consumption behaviors and identify any anomalies or outliers. This clustering process allows us to establish a reliable baseline of typical energy usage, which is crucial for developing effective DR strategies.

\subsection{Building's Available Flexibility Forecast Module}

The goal of this module is to accurately estimate the available flexibility in energy consumption for a building, based on historical data. Flexibility here refers to the ability to adjust energy consumption in response to external signals, such as DR events, without disrupting typical consumption patterns.

To achieve this, we begin by analyzing the building's energy consumption patterns over time. The primary tool we use for this analysis is the DFT, which converts time-domain data (e.g., hourly energy consumption) into the frequency domain. In the frequency domain, each frequency $f$ corresponds to a specific pattern or cycle in the data, such as daily or weekly usage patterns.

The frequency domain for DFT is defined by the sampling frequency $F_{s}$, which is calculated as:

\begin{equation} \label{eq0}
F_{s} = \frac{1}{T_{s}},
\end{equation}

\noindent where $T_{s}$ is the sampling interval, or the time between consecutive data points (e.g., one hour for hourly data). The sampling frequency $F_{s}$ represents how often the data is sampled per second.

DFT decomposes the time-series data into sinusoidal components, each with a specific frequency $f$. These frequencies correspond to recurring patterns in the energy consumption data. The magnitude of each frequency indicates how strongly that particular pattern is present in the data. For example, a dominant frequency might correspond to a 24-hour cycle, representing a daily usage pattern.

To make these frequencies more interpretable, we convert each frequency $f$ into a corresponding time period $t$ using the formula:

\begin{equation} \label{eq1}
t = \frac{1}{f}.
\end{equation}
 
Here, $t$ represents the time duration of the identified pattern (e.g., hours or days), and $f$ is the frequency obtained from the DFT analysis. This conversion helps us understand the actual time intervals over which the energy usage patterns repeat.

By focusing on the significant time periods $t$, we can effectively discern and anticipate habitual energy usage patterns, which are essential for developing and optimizing DR strategies. For instance, identifying a strong daily pattern (with $t=24$ hours) allows us to predict that certain consumption behaviors will repeat every day.

Rather than analyzing the data annually, we segment it into monthly intervals to capture more detailed trends and patterns. Each month’s data is further divided into smaller segments based on the time periods identified through DFT. For each segment, we calculate the minimum, medium, and maximum baseline values for energy consumption. These baselines define a range within which consumption levels typically fluctuate, providing bounds for estimating available flexibility without disrupting normal consumption patterns.

To refine these baselines, we apply the DBSCAN algorithm. DBSCAN helps cluster the data points by grouping similar energy consumption levels together and identifying outliers, such as peaks in consumption, that do not represent typical usage. The key parameters for DBSCAN are $\epsilon$ (the radius defining the neighborhood of a data point) and $k$ (the minimum number of points required to form a cluster). The k-Nearest Neighbors method is typically used to tune $\epsilon$, while k is manually selected.

We exclude clusters representing low energy consumption, as these typically indicate devices that cannot offer flexibility. Similarly, data points that show extreme high energy consumption, often appearing as peaks, are identified by DBSCAN as outliers and are also excluded from the analysis. The resulting clusters help us delineate more accurate minimum and maximum baselines. 

In the final step, the minimum and maximum baselines are adjusted based on the type of flexibility needed—upward for increasing energy consumption or downward for decreasing it. These adjustments, typically up to 10\% of the difference between the maximum and mean baseline, ensure the provided flexibility meets grid demands without disrupting normal operations. Aggregators can then use these estimates to offer flexibility to System Operators (SO) at specific intervals.

\section{Disaggregated Flexibility Provision}

The disaggregated flexibility provision strategy in this paper is centered on managing individual HVAC systems within a building, combining them into a Virtual Power Plant (VPP). In this context, a VPP refers to the aggregation of the flexibility from all HVAC units in a building, allowing them to function as a single, flexible entity capable of participating in DR events.

\subsection{HVAC's Available Flexibility Forecast Module}

Our methodology involves two models to estimate the flexibility of HVAC systems. The first model, the thermal model, predicts indoor temperature changes over the next time interval. The second model, the state predictor, uses the temperature predictions from the thermal model and the desired set temperature to forecast the HVAC's operational state (on/off). By combining the predicted states from the state predictor with the HVAC's rated power, we estimate the energy consumption for a given duration and temperature setting. Flexibility is then determined by calculating the difference in energy consumption between two different set temperatures, reflecting the HVAC's ability to adjust its power consumption in response to external signals while maintaining occupant comfort.

To prepare the data for model training, we down-sampled the data to accurately capture subtle temperature changes due to HVAC operation. We focused exclusively on periods when the HVAC was ON, ensuring that the models were trained with relevant, high-quality data. Key features, including indoor and outdoor temperatures, HVAC consumption, and operational state, were extracted and normalized to minimize the impact of scale differences. The dataset was then split into training and test sets. A linear regression model was used to predict indoor temperature changes (thermal model), while a random forest classifier was employed to forecast the next HVAC state (state predictor model).

Once the models are trained, they are integrated to estimate HVAC energy consumption over a specified duration and at a given set temperature. The process begins with the thermal model predicting the change in indoor temperature for the next time interval. Based on this temperature prediction and the chosen set temperature, the state predictor model determines whether the HVAC will be on or off in the subsequent time step. This process is repeated iteratively to predict the HVAC states over the entire duration of interest. Using these predicted states and the HVAC's rated power, the total energy consumption for the specified duration is calculated. The final step involves determining the available flexibility, which is defined as the difference in energy consumption between two distinct set temperatures, reflecting the HVAC's ability to adjust its power usage in response to external signals while maintaining occupant comfort. This process ultimately provides the predicted flexibility for each time step within the specified duration.

\subsection{VPP Optimal Configuration Module}

The goal of the VPP Optimal Configuration Module is to determine the optimal amount of flexibility each HVAC system can contribute to a DR event. Upon receiving a DR signal from the SO, the aggregator calculates the flexibility each customer can offer at each time step during the event. The available flexibility data from the HVAC's available flexibility forecast module, along with consumer contract and DR event details, inform this calculation. The contract details typically include the maximum flexibility each customer has agreed to provide, the terms of compensation, and any constraints related to comfort levels or operational limits. The DR event details provided by the SO include the duration of the event, the required lead time for notifications and the total demand reduction or increase needed. By integrating this information, the aggregator can ensure that the flexibility offered aligns with both the consumer’s contractual obligations and the SO’s requirements, optimizing the overall response to the DR event.

The objective function aims to minimize the difference between the requested DR power and the total flexibility provided by all occupants, which is the sum of the flexibilities of individual HVAC systems at each time step:

\begin{equation} \label{eq2}
\min_{\textbf{u}} \left( P_{DR} - \sum_{t=t_{s}}^{T} \sum_{i=1}^{N} P_{i, t}^{del} \right)^2,
\end{equation}

\noindent where \textbf{u} is the vector of decision variables $P_{i, t}^{del}$, namely the power delivered by the HVAC of occupant $i$ at timestep $t$. $P_{DR}$ is the amount that the aggregator requests for demand reduction in kW. $N$ is the total number of occupants and $t_{dur}$ is the entire duration of the DR event. The start of the DR event is $t_{s}$ and $T = t_{s} + t_{dur}$ is the end of the event. 

The flexibility provided by each HVAC system at each time step must not exceed the maximum power agreed upon in the contract:

\begin{equation} \label{eq3}
P_{i, t}^{del} \leq P_{con, i, t},
\end{equation}

\noindent where $P_{con, i, t}$ is the maximum amount of flexibility in kW that the occupant $i$ has agreed to provide at timestep $t$.

Finally, the provided flexibility by each occupant $i$ at each timestep $t$ should respect the available flexibility provided by the HVAC's available flexibility forecast module:

\begin{equation} \label{eq4}
P_{i, t}^{del} \leq P_{i, t}^{avail},
\end{equation}

\noindent where $P_{i, t}^{avail}$ is the available flexibility in kW for the occupant $i$ at timestep $t$ produced by the short-term available flexibility forecast for the HVAC.

\section{Results}

This section presents the outcomes of two case studies that reflect different data availability scenarios encountered by aggregators. The first case study deals with aggregated flexibility provision using data from a single smart meter monitoring total building consumption. The second case study focuses on disaggregated flexibility provision, utilizing data from dedicated metering devices attached to HVAC systems. Each scenario is supported by a dedicated web application designed to assist aggregators in making informed decisions.

\subsection{Aggregated Flexibility Provision}

\subsubsection{Case Study and Results}

We analyzed data from an industrial park in Cuerva, Spain, spanning from January 1, 2021, to December 31, 2021. The dataset includes hourly recordings of Active Imported Energy (AIE) from seven industrial buildings. 

Using the DFT analysis, we identified a prominent 24-hour cycle in the energy consumption pattern, reflecting the building's operational cycle. Figure \ref{fig:fig0} shows the hourly energy consumption for January, showing a consistent pattern with lower consumption during the night and higher, stable consumption during the day. Devices consuming 500 Wh or less were considered essential and not suitable for flexibility measures. We set a lower boundary of 500 Wh and an upper boundary of 2500 Wh to identify the range within which flexible devices operate.

\begin{center}
\includegraphics[width=0.8\columnwidth]{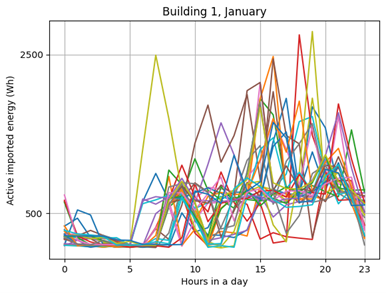}
\captionof{figure}{Hourly energy consumption of each day during the month of January for Building 1.}
\label{fig:fig0}
\end{center}

We applied the DBSCAN algorithm to cluster the data, which allowed us to identify typical consumption profiles while excluding outliers, as shown in Figure \ref{fig:fig1}. The DBSCAN algorithm was configured with $\epsilon = 0.5$ kWh and $k = 5$, achieving a Silhouette Coefficient of 0.72, indicative of well-separated clusters.

\begin{center}
\includegraphics[width=0.8\columnwidth]{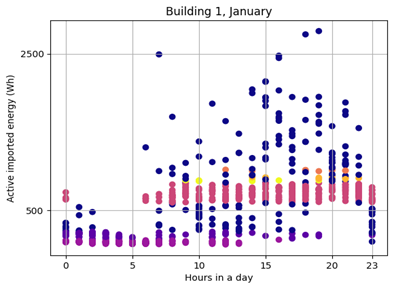}
\captionof{figure}{Clustering of hourly energy consumption using the DBSCAN algorithm.}
\label{fig:fig1}
\end{center}

Based on these clusters, we established minimum, medium, and maximum baselines, as depicted in Figure \ref{fig:fig2}. These baselines serve as reference points for understanding energy consumption patterns and assessing flexibility potential.

\begin{center}
\includegraphics[width=0.8\columnwidth]{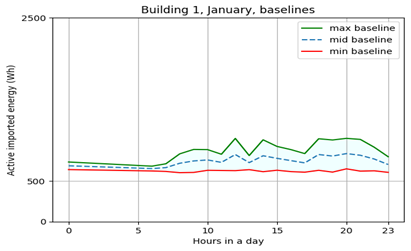}
\captionof{figure}{Energy consumption baselines for assessing flexibility potential.}
\label{fig:fig2}
\end{center}

\subsubsection{Web App for Aggregators}

The derived baselines are crucial for responding to flexibility requests. We developed a web application to provide aggregators with essential insights into flexibility availability. The application offers a user-friendly interface where aggregators can select assets, view upcoming flexibility requests, and configure VPP settings using an interactive slider. This tool enables aggregators to respond effectively to flexibility requests, enhancing their participation in the energy market.

\begin{center}
\includegraphics[width=0.8\columnwidth]{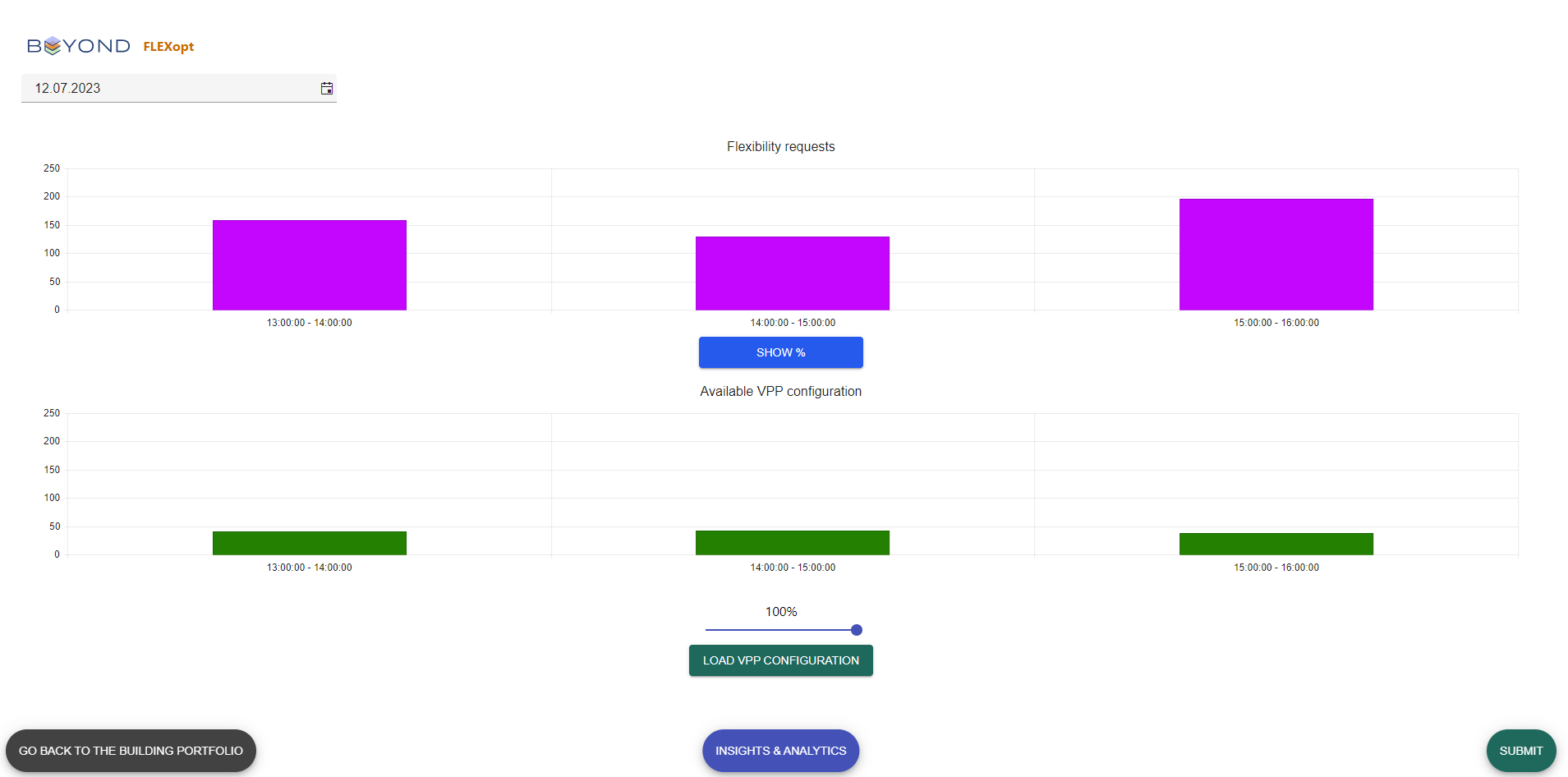}
\captionof{figure}{Interface of the web application showing the available VPP configuration options for aggregators.}
\label{fig:fig3}
\end{center}

\subsection{Disaggregated Flexibility Provision}

\subsubsection{Case Study and Results}

This case study examines a residential building in Madrid, Spain, comprising eight apartments equipped with dedicated HVAC systems and metering devices. The primary objective is to assess the potential of these HVAC systems in providing flexibility  during DR events triggered by an aggregator.

We trained a short-term flexibility forecast model for each apartment, considering HVAC usage patterns and environmental conditions. The thermal model achieved a Mean Absolute Error (MAE) of $0.3^{\circ}$ C and an $R^{2}$ score of 0.88 on the test set and the state predictor achieved an accuracy of 94.2\% and an F1-Score of 93.2\%. Figure \ref{fig:fig4} illustrates the predicted flexibility for the HVAC system of apartment 4 over a three-hour period at 15-minute intervals (12 steps in total).

\begin{center}
\includegraphics[width=0.8\columnwidth]{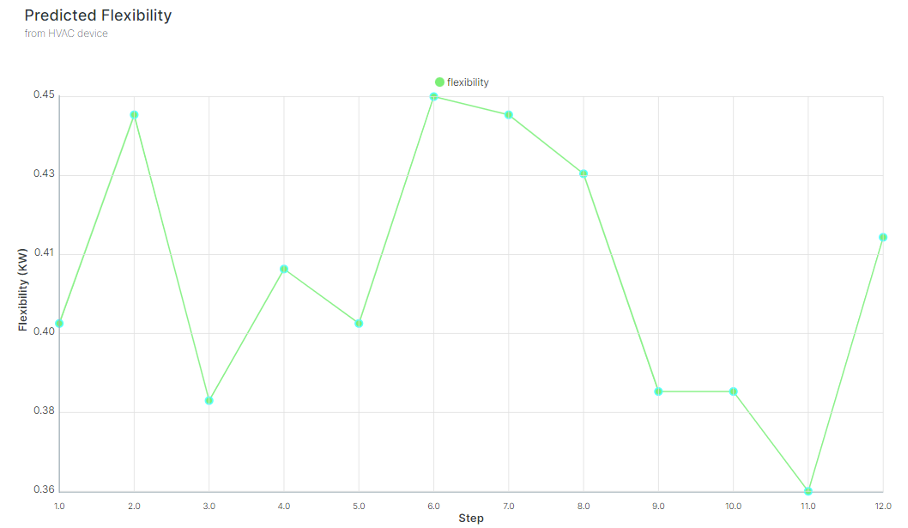}
\captionof{figure}{3-hour ahead flexibility prediction for HVAC system of apartment 4.}
\label{fig:fig4}
\end{center}

Upon receiving a DR signal, the VPP optimal configuration module was employed to compute the optimal flexibility contribution from each HVAC system, ensuring that the flexibility provision adhered to contract details and the available flexibility forecasts. 

\subsubsection{Web App for Aggregators}

The developed web application provides aggregators with essential tools to manage and optimize the participation in DR events in real-time. When a flexibility event is triggered by the SO, the VPP optimizal configuration problem is automatically solved, and the details of the DR event and the solution are displayed user interface. Figure \ref{fig:fig5} shows the tab that allows aggregators to review and analyze the specific contributions requested from each HVAC system.

\begin{center}
\includegraphics[width=0.8\columnwidth]{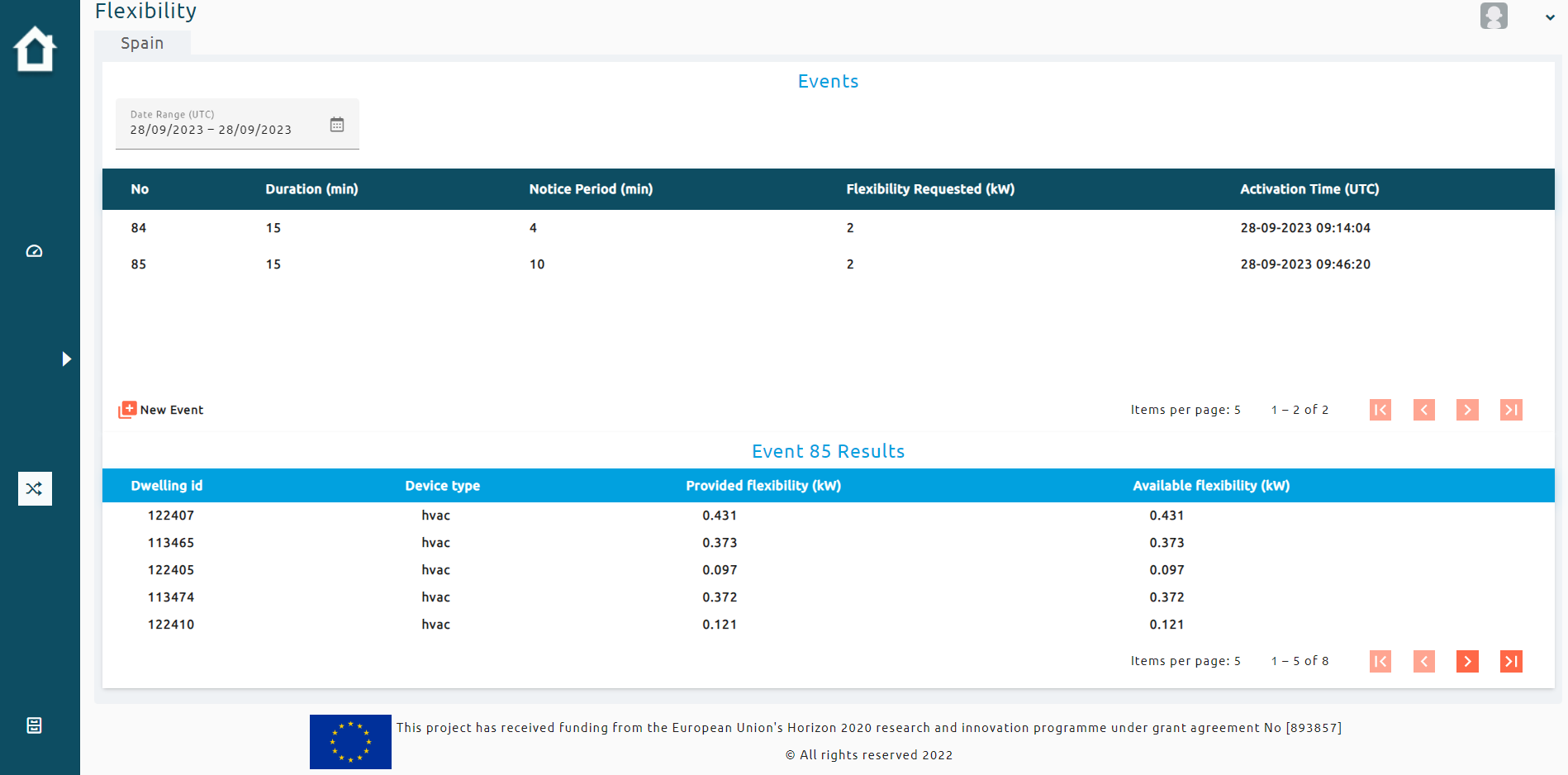}
\captionof{figure}{Detailed breakdown of DR event contributions from each HVAC system.}
\label{fig:fig5}
\end{center}

Additionally, the user interface features an overview tab that enables aggregators to monitor the fulfillment of flexibility requests. This tab provides a comparison between the initially requested flexibility and the actual flexibility delivered by the devices, as depicted in Figure \ref{fig:fig6}. This functionality is crucial for evaluating the performance and reliability of the HVAC systems participating in the DR events, offering valuable insights to help aggregators optimize their demand response strategies.

\begin{center}
\includegraphics[width=0.8\columnwidth]{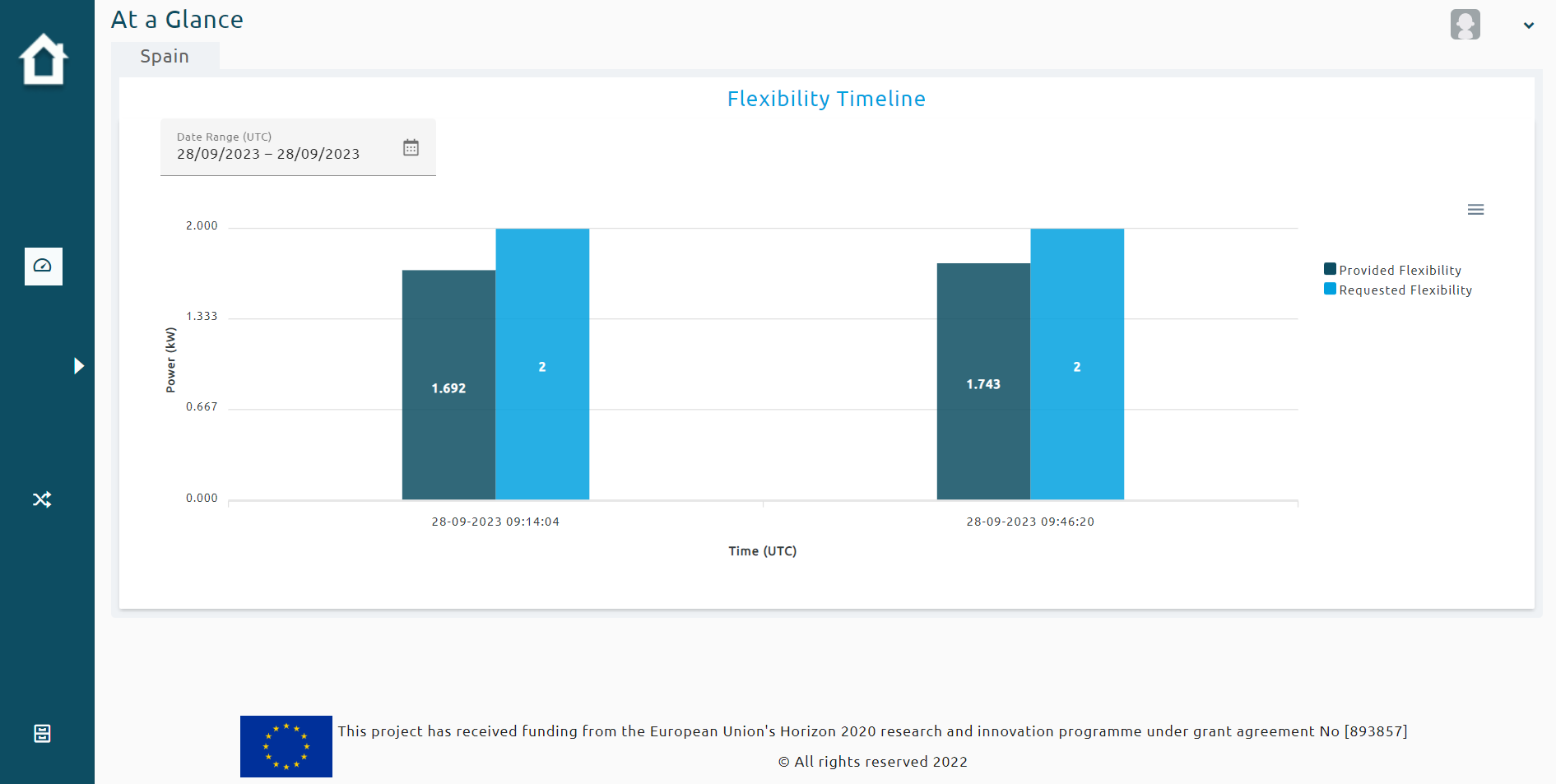}
\captionof{figure}{Overview of requested versus actual flexibility contributions.}
\label{fig:fig6}
\end{center}

\section{Conclusions and Discussion}

This paper presents a comprehensive and practical approach to flexibility provision strategies, focusing on both aggregated and disaggregated methods tailored to the specific data available to aggregators. Our approach is distinguished by its emphasis on creating user-friendly and easily understandable tools, which are crucial for aggregators seeking to effectively participate in DR events. The case studies underscore the real-world applicability of these methods in optimizing energy consumption and maximizing flexibility across diverse scenarios.

A key strength of this work lies in its dual focus: not only does it provide analytical tools, but it also prioritizes consumer acceptance, ensuring that the strategies proposed are not just theoretically sound but also practically viable. By incorporating thermal comfort into the disaggregated flexibility module, the approach aligns with the needs and preferences of consumers, thereby enhancing engagement and participation rates in DR programs.

The integration of aggregated and disaggregated approaches, supported by the appropriate infrastructure and data collection practices, offers significant potential for advancing research and innovation in the energy sector. As the energy landscape evolves towards greater decarbonization and increased integration of renewable energy sources, the need for flexibility will become increasingly critical. The methodologies and insights provided in this study not only address current challenges but also establish a robust framework for future advancements in energy flexibility provision and management.

\section{Acknowledgments}

This work was supported by the BEYOND, frESCO and SYNERGIES projects. The BEYOND project has received funding from the European Union’s Horizon 2020 Research and Innovation program under No.957020, frESCO has received funding from the European Union’s Horizon 2020 Research and Innovation program under No.893857 and SYNERGIES has received funding from the European Union’s Horizon Europe program, Grant Agreement No.101069839.


%


\bibliographystyle{unsrt}
\bibliography{bibliography}

\end{multicols}

\end{document}